\documentclass[10pt,journal,a4paper]{IEEEtran}%TWC
\usepackage{amssymb}
\usepackage{amsmath}
\usepackage{bm}
\usepackage{amsthm}
\usepackage{graphicx}
\usepackage{subfigure}
\usepackage{float}
\usepackage{cite}
\usepackage{enumerate}
\usepackage{enumitem}
\usepackage{multirow}
\usepackage{amsfonts}
\usepackage{url}
\usepackage{diagbox}
\usepackage{color, soul}
\usepackage{amsthm}
\usepackage{type1cm}

\newcommand{\tabincell}[2]{\begin{tabular}{@{}#1@{}}#2\end{tabular}}
\newcommand{\titlefont}{\fontsize{16.5pt}{24pt}\selectfont}

\begin{document}
\newtheorem{remark}{\bf~~Remark}
\newtheorem{proposition}{\bf~~Proposition}
\newtheorem{theorem}{\bf~~Theorem}
\newtheorem{definition}{\bf~~Definition}
%\font\myfont=cmr12 at 14pt
\title{{\titlefont Reconfigurable Intelligent Surfaces in 6G: Reflective, Transmissive, or Both?}}

\author{
\IEEEauthorblockN{
	{Shuhao Zeng}, \IEEEmembership{Student Member, IEEE},
	{Hongliang Zhang}, \IEEEmembership{Member, IEEE},
	{Boya Di}, \IEEEmembership{Member, IEEE},\\
	{Yunhua Tan}, \IEEEmembership{Member, IEEE},
	{Zhu Han}, \IEEEmembership{Fellow, IEEE},
	{H. Vincent Poor}, \IEEEmembership{Life Fellow, IEEE},\\
	{and Lingyang Song}, \IEEEmembership{Fellow, IEEE}}
	\vspace{-0.9cm}
	\thanks{S. Zeng, Y. Tan, and L. Song are with Department of Electronics, Peking University, Beijing, China (email: \{shuhao.zeng, tanggeric,  lingyang.song\}@pku.edu.cn).}
	\thanks{H. Zhang and H. V. Poor are with Department of Electrical Engineering, Princeton University, USA (email: hongliang.zhang92@gmail.com, poor@princeton.edu).}
	\thanks{B. Di is with Department of Computing, Imperial College London, UK (email: diboya92@gmail.com).}
	\thanks{Z. Han is with Electrical and Computer Engineering Department, University of Houston, Houston, TX, USA, and also with the Department of Computer Science and Engineering, Kyung Hee University, Seoul, South Korea (email: zhan2@uh.edu).}
}

\maketitle

\begin{abstract}
Reconfigurable intelligent surfaces~(RISs) have attracted wide interest from industry and academia since they can shape the wireless environment into a desirable form with a low cost. In practice, RISs have three types of implementations: 1) reflective, where signals can be reflected to the users on the same side of the base station~(BS), 2) transmissive, where signals can penetrate the RIS to serve the users on the opposite side of the BS, and 3) hybrid, where the RISs have a dual function of reflection and transmission. However, existing works focus on the reflective type RISs, and the other two types of RISs are not well investigated. In this letter, a downlink multi-user RIS-assisted communication network is considered, where the RIS can be one of these types. We derive the system sum-rate, and discuss which type can yield the best performance under a specific user distribution. Numerical results verify our analysis.

%Since the sum rate of the network is significantly influenced by the operation mode of the RIS, we investigate about the optimal mode that maximizes the sum rate, which is challenging since the expression of the sum rate is complicated. To cope with the challenge, we approximate the network geometry and utilize the law of large number, and the optimal modes in different cases are discussed. Numerical results verify our analysis.
\end{abstract}
\vspace{-0.1cm}
\vspace{-0.45cm}
\section{Introduction}
%\vspace{-0.1cm}
The recent development of meta-surfaces has given rise to a new technology, reconfigurable intelligent surfaces~(RISs), which can control the wireless communication environment and improve the signal quality at the receiver~\cite{MHLKZG_2020}. The RIS is an ultra-thin surface inlaid with many sub-wavelength units, whose electromagnetic~(EM) response can be tuned via onboard positive-intrinsic-negative~(PIN) diodes. According to the energy split for reflection and transmission, the RISs can be roughly categorized into three types: reflective, transmissive~\cite{YLBWRXQT}, and hybrid~\cite{SHBYZL_2020,XJB_2018}. Specifically, the reflective type RIS reflects incident signals towards the users on the same side of the base station~(BS) while signals can penetrate the transmissive type RIS towards the users on the opposite side. For the hybrid type, the RIS enables a dual function of reflection and transmission.

In the literature, current researches focus on the reflective type RIS. For example, in~\cite{BR_2020}, a point-to-point orthogonal frequency division multiplexing network assisted by a reflective type RIS was studied, where the authors proposed a practical transmission protocol to estimate channel and optimize reflection successively. In~\cite{BCR_2020}, two efficient channel estimation schemes were proposed for different channel setups in a reflective type RIS-assisted multi-user network using orthogonal frequency division multiple access, where the pilot tone allocations and RIS configurations were further optimized to minimize the channel estimation error. In~\cite{BHLYZH_2020}, a multi-user network assisted by a reflective type RIS was investigated, where the digital beamformer at the BS and RIS configurations were jointly optimized to maximize the sum rate.

In this letter, we consider an RIS-assisted downlink multi-user network with one BS, where the RIS can be any one of the three types. We first derive an upper bound of the system sum-rate. Then, based on the upper bound, we discuss the optimal type of the RIS to be deployed under different UE distributions. Finally, simulation results verify the effectiveness of our derivation and the proposed optimality conditions.
\vspace{-0.7cm}
\section{System Model}%
\vspace{-0.2cm}
\label{sys_mod}
%\vspace{-0.1cm}
%In this section, we first introduce the system model of RIS-assisted networks. Afterwards, characteristics of the RIS are discussed, and the channel model is presented.
%\vspace{-0.3cm}
\subsection{Scenario Description}
\vspace{-0.1cm}
\begin{figure}[!tpb]
	\centering
	\includegraphics[width=2.8in]{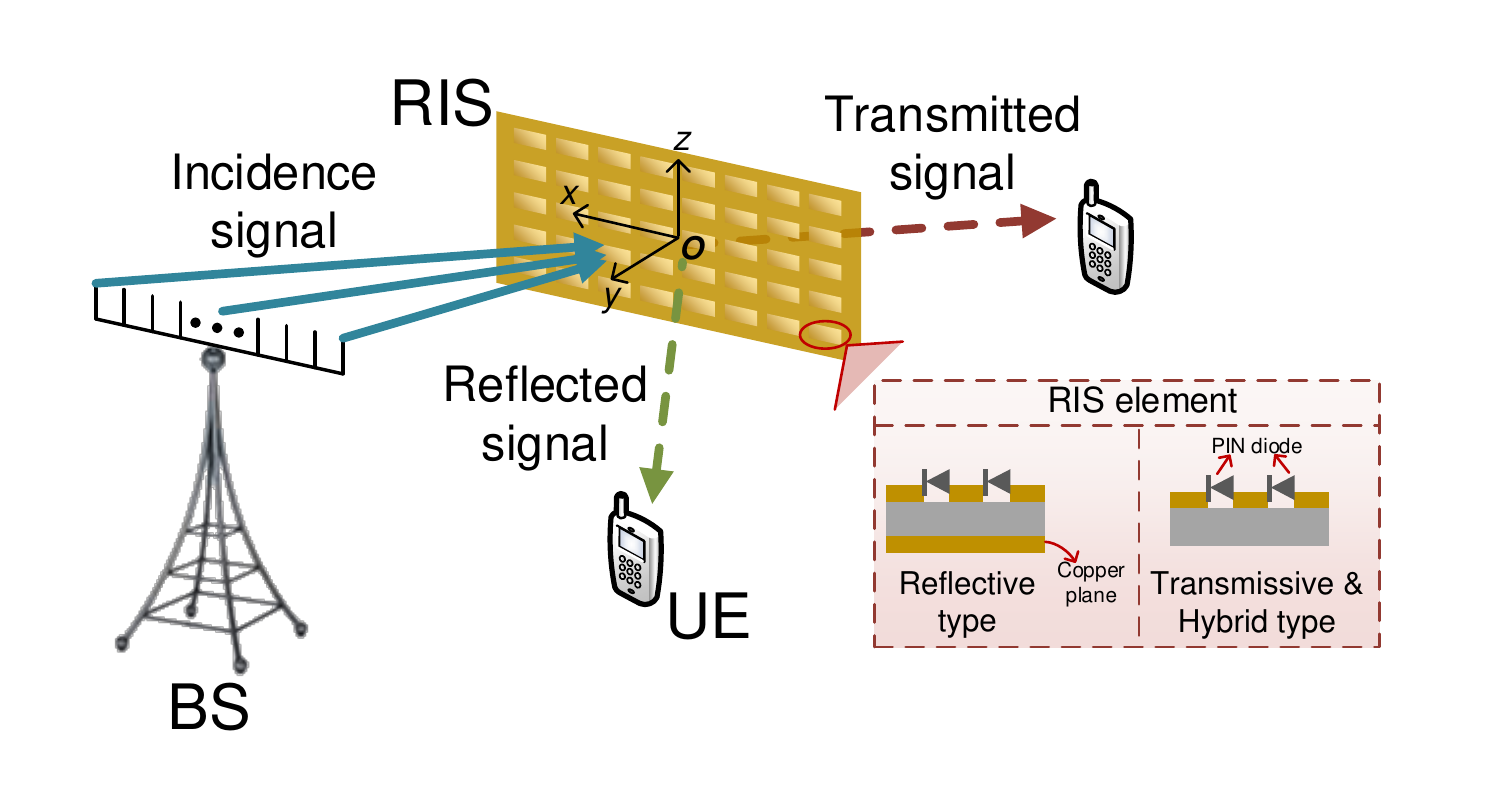}
	\vspace{-0.6cm}
	\caption{System model of an RIS-assisted cellular network.}
	\vspace{-0.7cm}
	\label{system_model}
\end{figure}
As shown in Fig.~\ref{system_model}, we consider a downlink multi-user communication system, consisting of a BS equipped with $K_t$ antennas and $S$ UEs with one single antenna. Due to the complicated and dynamic wireless environment involving unexpected fading and potential obstacles, the links from the BS to the UEs may be unstable or even fall into complete outage. To alleviate this issue, we deploy an RIS to assist the communication, which reflects or penetrates the signals from the BS towards the UEs by shaping the propagation environment into a desirable form. To describe the topology of the network, we introduce a Cartesian coordinate, where the $xoz$ plane coincides with the RIS, with the origin at the center of the RIS. For simplicity of explanation, regions $y<0$ and $y>0$ are referred to as \emph{transmission zone} and \emph{reflection zone}, respectively. We use $\mathcal{S}_R$ and $\mathcal{S}_T$ to represent the set of UEs which are located at the reflection zone and transmission zone, respectively, with the number of such UEs denoted by $S_R$ and $S_T$, respectively. Therefore, we have $S=S_T+S_R$.
\vspace{-0.4cm}
\subsection{Reconfigurable Intelligent Surfaces}
An RIS is composed of $M \times N$ sub-wavelength elements, each with the size of $l_M\times l_N$. As shown in Fig.~\ref{system_model}, the element has several positive-intrinsic-negative~(PIN) diodes onboard. Define state of an element as the biased voltage applied to the PIN diodes on the elements. When the state of an element becomes different, the electromagnetic response of this element, i.e., phase shift, varies accordingly. 

Denote the amplitude for the transmission and reflection responses of one element by $\Gamma^t$ and $\Gamma^r$, respectively\footnote{We assume that the RIS elements are uniform, and thus, have the same amplitude response.}. For simplicity, we assume that no power is dissipated by the RIS elements, i.e., $(\Gamma^t)^2+(\Gamma^r)^2=1$. To meet diverse requirements in 6G, three types of RISs are considered here: reflective, transmissive, and hybrid types. In the following, we  will elaborate on the three types.
\begin{enumerate}[itemindent=0em, label=$\bullet$]
	\item \textbf{Reflective type}: Each element only reflects the incident signals due to the copper backplane. Therefore, we have $\Gamma^t=0$ and $\Gamma^r=1$ for the reflective type RIS. 
	\item \textbf{Transmissive type}: Without the copper backplane, the incident signals can only penetrate the elements, i.e., $\Gamma^r=0$ and $\Gamma^t=1$.
%	\footnote{\color{blue}The penetration loss caused by the transmissive type is similar to the reflection loss of the reflective type RIS.}
	\item \textbf{Hybrid type}: Similar to the transmissive type, the copper backplane is removed. However, by adjusting RIS structures such as element size, and metallic patch patterns, the incurred currents within the RIS can be different, which thus leads to both the reflection and transmission of the incident signals, i.e., $\Gamma^r>0$ and $\Gamma^t>0$. In this letter, we assume that the energy is equally split for transmission and reflection, i.e., $(\Gamma^r)^2 =(\Gamma^t)^2=\frac{1}{2}$, which is the most typical implementation of the hybrid type RISs. 
\end{enumerate}

Define $\varphi^r$ and $\varphi^t$ as the reflection and transmission phase shift of one element, respectively. By combining the amplitudes and phase shifts, the reflection and transmission coefficients of the $(m,n)$-th element can be written as $\Gamma_{m,n}^r=\Gamma^re^{-j\varphi^r_{m,n}}$ and $\Gamma_{m,n}^t=\Gamma^te^{-j\varphi^t_{m,n}}$, respectively.    

%\footnote{\color{blue}The analysis in this paper does not rely on the assumption that the transmission and reflection phases of the hybrid type RIS are coupled.}

%Note that $\varphi^r$ and $\varphi^t$ are coupled?. Therefore, we assume that $\varphi^r=\varphi^t$ without loss of generality.
\vspace{-0.4cm}
\subsection{Channel Model}
\label{channel_model}
Define $\bm{H}$ as the channel matrix between the BS and UEs, where $[\bm{H}]_{s,k}\triangleq h^{(s,k)}$ is the channel gain from the $k$-th antenna of the BS to UE $s$. The channel from antenna $k$ of the BS to UE $s$ is composed of $MN$ RIS-based channels, where the direct link is assumed to be blocked. Besides, among the $MN$ RIS-based channels, the $(m,n)$-th channel represents the channel from antenna $k$ to UE $s$ via the $(m,n)$-th RIS element.

The RIS-based channels are modeled by considering fast fading, path loss, and RIS responses. Specifically, the $(m,n)$-th RIS-based channel can be written as
{\setlength\abovedisplayskip{0.1cm}
	\setlength\belowdisplayskip{0.05cm}
\begin{align}
\label{RIS_based}
\hat{h}_{m,n}^{(s,k)}=\sqrt{\beta_{m,n}^{(s,k)}}g_{m,n}^{(s,k)}\Gamma_{m,n}^{(s)},
\end{align}
where $\beta_{m,n}^{(s,k)}$ represents pathloss, $g_{m,n}^{(s,k)}$ is small scale fading coefficient with the mean being $0$ and the variance being $1$, which are assumed to be independent for different RIS-based channels\footnote{The distribution of the small scale fading can be arbitrary.}, and $\Gamma_{m,n}^{(s)}$ represents the RIS element response, where $\Gamma_{m,n}^{(s)}=\Gamma_{m,n}^r$ when the BS and UE $s$ are on the same side of the RIS, and $\Gamma_{m,n}^{(s)}=\Gamma_{m,n}^t$ when they are separated by the RIS. Based on~\cite{WMXJYMYSQT}, the pathloss can be modeled by
\begin{align}
\beta_{m,n}^{(s,k)}=\frac{\lambda^2Gl_Ml_N}{64\pi^3}\frac{G_{I}F_{m,n}^{(k)}F_{m,n}^{(s)}}{(D_{m,n}^{(k)}d_{m,n}^{(s)})^{\alpha}},
\end{align}
where $\lambda$ is the wavelength corresponding to the carrier frequency, $G$ is antenna gain, $\alpha$ represents the pathloss exponent, $F_{m,n}^{(k)}$ is the normalized power radiation pattern of the $(m,n)$-th RIS element for the arrival signal from the $k$-th BS antenna, $F_{m,n}^{(s)}$ is the normalized power radiation pattern of the $(m,n)$-th RIS element for the departure signal to UE $s$, $G_I$ is the antenna gain of one RIS element, $D_{m,n}^{(k)}$ and $d_{m,n}^{(s)}$ are the distance between antenna $k$ and the $(m,n)$-th RIS element, and the distance between the $(m,n)$-th RIS element and UE $s$, respectively. According to~\cite{OEE_2020}, the departure angle has no influence on one RIS-based channel while the effect of the incident angle can be modeled by $\cos\theta_{m,n}^{(k)}$, where $\theta_{m,n}^{(k)}$ is the angle between the incident signal and the $y$-axis. Therefore, $F_{m,n}^{(k)}$ and $F_{m,n}^{(s)}$ are modeled by}
{\setlength\abovedisplayskip{0cm}
	\setlength\belowdisplayskip{0cm}
\begin{align}
F_{m,n}^{(k)}=\cos^2\theta_{m,n}^{(k)},
\end{align}
\begin{equation}
F_{m,n}^{(s)}=\left\{
\begin{array}{lr}
\epsilon_r,~\text{UE $s$ is in the reflection zone},\\
\epsilon_t,~\text{otherwise},
\end{array}
\right.
\end{equation}
%\begin{salign}
%	F_{m,n}^{(s)}=\left\{
%	\begin{aligned}
%		&\epsilon_r,~\text{UE $s$ is in $y>0$},\\
%		&\epsilon_t,~\text{otherwise},
%	\end{aligned}
%	\right.
%\end{salign}
respectively, where $\epsilon_r$ and $\epsilon_t$ are constants related to the hardware implementation.}

By combining these RIS-based channels, the channel from antenna $k$ of the BS to UE $s$ can be written as
{\setlength\abovedisplayskip{0.1cm}
	\setlength\belowdisplayskip{-0.1cm}
\begin{align}
\label{ch_IS}
h^{(s,k)}=\sum_{m,n}\hat{h}_{m,n}^{(s,k)}=\sqrt{\hat{\beta}^{(s)}}\sum_{m,n}g_{m,n}^{(s,k)}\Gamma_{m,n}^{(s)},
\end{align}
where} $\hat{\beta}^{(s)}$ is average pathloss between UE $s$ and the BS\footnote{We assume that the distance between the BS and the RIS is much larger than the size of the antenna array at the BS and the size of the RIS. Besides, the distance between the BS and UE $s$ is also assumed to be much larger than the size of the RIS. Therefore, the pathloss between UE $s$ and different antennas via different RIS elements can be regarded as the same.}.
\vspace{-0.2cm}
\section{Analysis on Sum Rate}
\vspace{-0.1cm}
\label{rate_ana}
According to~\cite{DP_2005}, the additive white Gaussian noise~(AWGN) channel capacity can be achieved by dirty paper coding, and the AWGN capacity is given by
{\setlength\abovedisplayskip{0.1cm}
	\setlength\belowdisplayskip{0.05cm}
\begin{align}
\label{cap}
C=\mathbb{E}\Big(\sum_s\log_2(1+\frac{P_T}{\sigma^2}\Lambda_s[\bm{H}\bm{H}^{\mathrm{H}}]_{s,s})\Big),
\end{align}
where $P_T$ is transmit power of the BS, $\sigma^2$ represents noise variance, and $\Lambda_s$ is $s$-th element of power allocation vector $\bm{\Lambda}$, with $\sum_{s}\Lambda_s=1$.}

In the following, an upper bound on the system capacity will be presented. Before presenting the upper bound, we first give a proposition on the channel.
\vspace{-0.1cm}
\begin{proposition}
	\label{gain_dis}
	When number $MN$ of RIS elements is large, and the phase shifts $\varphi_{m,n}^t$ and $\varphi_{m,n}^r$ of the RIS elements are deterministic, we can derive that
	{\setlength\abovedisplayskip{0.05cm}
		\setlength\belowdisplayskip{-0.2cm}
%		{\footnote{\color{blue}The phase response of the RIS elements can be different, which can be adjusted according to network settings.}}
	\begin{align}
	\Big(\sum_{m,n}g_{m,n}^{(s,k)}\Gamma_{m,n}^{(s)}\Big)/\sqrt{MN}\sim \mathcal{CN}(0,|\Gamma_{m,n}^{(s)}|^2). 
	\end{align}}
\end{proposition}
\vspace{-0.3cm}
\begin{proof}
	Note that the small scale fading for different RIS elements are independently identically distributed~(i.i.d.). Therefore, when the phase shifts of the RIS elements are given, according to the central limit theorem, the channel gain corresponding to one link from the BS antennas to the UEs follows a complex normal distribution when the number of RIS elements is large.
	
	Since the mean for the small scale fading is $0$, the mean of the Gaussian distribution is equal to $0$. Besides, the variance for the small scale fading is $1$ and the amplitude of the RIS response is $|\Gamma_{m,n}^{(s)}|$. Therefore, the variance of the Gaussian distribution is $|\Gamma_{m,n}^{(s)}|^2$. 
\end{proof}
\vspace{-0.3cm}
	Based on the proposition, we can derive an upper bound for the system capacity, as shown in the following proposition.
	\vspace{-0.1cm}
\begin{proposition}
	\label{prop_cap_ub}
	The capacity can be upper bounded by
	{\setlength\abovedisplayskip{0.1cm}
		\setlength\belowdisplayskip{0cm}
	\begin{align}
	\label{up_capacity}
	C\le \sum_s\log_2\left(1+\frac{P_T}{\sigma^2}\Lambda_s\hat{\beta}^{(s)}K_tMN|\Gamma_{m,n}^{(s)}|^2\right)\triangleq C^{ub}.
	\end{align}}
\end{proposition}
\vspace{-0.4cm}
\begin{proof}
	According to Jensen's inequality, we have
	{\setlength\abovedisplayskip{0.1cm}
		\setlength\belowdisplayskip{0cm} 
	\begin{align}
	\label{cap_ub}
	C&\le \sum_s\log_2\Big(1+\frac{P_T}{\sigma^2}\Lambda_s\mathbb{E}([\bm{H}\bm{H}^{\mathrm{H}}]_{s,s})\Big)\notag\\
	&=\sum_s\log_2\Big(1+\frac{P_T}{\sigma^2}\Lambda_s\mathbb{E}(\sum_k|h^{(s,k)}|^2)\Big).	
%	\sum_s\log_2(\mathbb{E}(1+\frac{\gamma_0\lambda_s}{S}[\bm{H}\bm{H}^{\mathrm{H}}]_{s,s}))
	\end{align}
	Based on the expression of channel gain in (\ref{ch_IS}), we have
%	\begin{align}
%	\label{10_step_1}
%	\mathbb{E}\Big(\sum_k|h^{(s,k)}|^2\Big)&=\sum_k\mathbb{E}\left(|h^{(s,k)}|^2\right)\notag\\
%	&=\sum_k\hat{\beta}^{(s)}\mathbb{E}\Big(\Big|\sum_{m,n}g_{m,n}^{(s,k)}\Gamma_{m,n}^{(s)}\Big|^2\Big).
%	\end{align}
\begin{align}
\label{10_step_1}
\mathbb{E}\Big(\sum_k|h^{(s,k)}|^2\Big)=\sum_k\hat{\beta}^{(s)}\mathbb{E}\Big(\Big|\sum_{m,n}g_{m,n}^{(s,k)}\Gamma_{m,n}^{(s)}\Big|^2\Big).
\end{align}
%	\begin{align}
%	\label{10_step_1}
%	\mathbb{E}\Big(\!\sum_k\!|h^{(s,k)}|^2\!\Big)\!=\!\sum_k\!\mathbb{E}\left(\!|\!h^{(s,k)}\!|^2\!\right)\!=\!\sum_k\!\hat{\beta}^{(s)}\mathbb{E}\Big(\!\Big|\!\sum_{m,n}g_{m,n}^{(s,k)}\Gamma_{m,n}^{(s)}\!\Big|^2\!\Big).
%	\end{align}
	Besides, according to Proposition 1, it can be obtained that
	\begin{align}
	\label{10_step_2}
	\mathbb{E}\Big(\Big|\sum_{m,n}g_{m,n}^{(s,k)}\Gamma_{m,n}^{(s)}\Big|^2\Big)=MN\Big|\Gamma_{m,n}^{(s)}\Big|^2.
	\end{align}
	By combining (\ref{10_step_1}) and (\ref{10_step_2}), we can derive that
	\begin{align}
	\mathbb{E}\Big(\sum_k|h^{(s,k)}|^2\Big)=\hat{\beta}^{(s)}K_tMN|\Gamma_{m,n}^{(s)}|^2,
	\end{align}
	which ends the proof.}
\end{proof}
\vspace{-0.2cm}
Denote the conjugate transpose operator by $(\cdot)^{\mathbb{H}}$. The derived upper bound is shown to be tight in the following remark.
\vspace{-0.5cm}
\begin{remark}
	Upper bound $C^{ub}$ can be achieved when the eigenvalues of $\bm{HH^{\mathbb{H}}}$ are the same, which can be obtained by optimizing the phase shifts to maximize the effective rank of $\bm{H}$~\cite{how_many_ele}.
\end{remark}
\vspace{-0.3cm}
In the following, the upper bound $C^{ub}$ given in Proposition~\ref{prop_cap_ub} is selected to evaluate the performance of different RIS types. Since the distances between the UEs and the RIS will influence the system performance, the distances between the UEs~(in both $\mathcal{S}_T$ and $\mathcal{S}_R$) and the RIS are assumed to be the same for simplicity\footnote{The analysis of the optimal type when the distances between the UEs and the RIS are different is left as future work.}. Moreover, since we evaluate the long-term performance, we consider the pathloss instead of the instantaneous channel information when allocating the power, i.e., the small scale fading is neglected here. Under these assumptions, the power will be allocated in the following way. 

\textbf{For the transmissive (reflective) type}, on the one hand, the UEs in $\mathcal{S}_R$ ($\mathcal{S}_T$) will not be allocated power to maximize the system performance. On the other hand, since the data rate is a concave function with respect to the transmit power, an equal power allocation among the UEs subjected to a total power budget will lead to the maximum system capacity. Therefore, the transmit power for UEs in $\mathcal{S}_T$ ($\mathcal{S}_R$) should be the same. 

\textbf{For the hybrid type}, the transmit power is allocated equally to UEs on the same side of the IOS. Define $\Lambda_R$ and $\Lambda_T$ as the power allocations for UEs in $\mathcal{S}_R$ and $\mathcal{S}_T$, respectively, i.e., $\Lambda_s=\Lambda_R, s\in\mathcal{S}_R$, and $\Lambda_s=\Lambda_T, s\in \mathcal{S}_T$. Since the power allocation is constrained by $\sum_s\Lambda_s=1$, we have $\Lambda_T=\frac{1-S_R\Lambda_R}{S_T}$. Therefore, the system capacity $C^{ub}$ is a function of one single variable $\Lambda_R$, where the optimal power allocation can be expressed as
	{\setlength\abovedisplayskip{0cm}
		\setlength\belowdisplayskip{0.05cm}
	\begin{equation}
	\Lambda_s=\left\{
	\begin{array}{lr}
	\Lambda^*,~\text{UE $s$ is in $\mathcal{S}_R$},\\
	\frac{1-S_R\Lambda^*}{S_T},~\text{otherwise}.
	\end{array}
	\right.
	\end{equation}
	Here,  $\Lambda^*=\max\Big(0,\min\Big(\frac{1}{S_R},L\frac{2S_T}{S}(\frac{1}{\epsilon_t}-\frac{1}{\epsilon_r})+\frac{1}{S}\Big)\Big)$. The expression of $L$ is given by $L=64\pi^3(D_{c}^{(c)}d_{c,U})^{\alpha}\sigma^2/(P_T\lambda^2Gl_Ml_NG_{I}\cos^2\theta_c^{(c)}K_tMN)$,
%	{\setlength\abovedisplayskip{0cm}
%		\setlength\belowdisplayskip{0.05cm}
%	\begin{align}
%	L=\frac{64\pi^3(D_{c}^{(c)}d_{c,U})^{\alpha}}{\gamma_0\lambda^2Gl_Ml_NG_{I}\cos^2\theta_c^{(c)}K_tMN},
%	\end{align}
	where $D_c^{(c)}$ is the distance between the BS antenna array center and the RIS center, $d_{c,U}$ denotes the distance between the RIS center and one UE, and $\theta_c^{(c)}$ is the angle between the $y$-axis and an incident signal from the BS antenna array center.}
 
With the above power allocation scheme, the sum-rate $C^{ub}$ with the reflective type, transmissive type, and hybrid type RIS can be expressed as
%\begin{center}
%\begin{align}
{\setlength\abovedisplayskip{0cm}
	\setlength\belowdisplayskip{0cm}
%\begin{gather}
%C_R=S_R\log_2(1+\frac{\epsilon_r}{L}\frac{1}{S_R}),\\
%C_T=S_T\log_2(1+\frac{\epsilon_t}{L}\frac{1}{S_T}),
%\end{gather}}
%\begin{gather}
%C_R=S_R\log_2\left(1+\frac{\epsilon_r}{L}\frac{1}{S_R}\right),
%C_T=S_T\log_2\left(1+\frac{\epsilon_t}{L}\frac{1}{S_T}\right),
%\end{gather}}
\begin{gather}
C_R\!=\!S_R\log_2\left(\!1\!+\!\frac{\epsilon_r}{LS_R}\!\right),
C_T\!=\!S_T\log_2\left(\!1\!+\!\frac{\epsilon_t}{LS_T}\!\right),
\end{gather}
{\setlength\abovedisplayskip{0.1cm}
	\setlength\belowdisplayskip{0cm}
%\begin{align}
%C_H=&S_T\log_2\left(1\!+\!\frac{\epsilon_t}{2L}\frac{1\!-\!S_R\Lambda^*}{S_T}\right)\notag\\
%&+\!S_R\log_2\left(1\!+\!\frac{\epsilon_r\Lambda^*}{2L}\right),
%\end{align}
\begin{align}
C_H=&S_T\log_2\left(1\!+\!\frac{\epsilon_t}{2L}\frac{1\!-\!S_R\Lambda^*}{S_T}\right)+\!S_R\log_2\left(1\!+\!\frac{\epsilon_r\Lambda^*}{2L}\right).
\end{align}}
%\end{align}
%\end{center} 
%\vspace{-0.2cm}
%\begin{align}
%C_T=S_T\log_2(1+\frac{\epsilon_t}{L}\frac{1}{S_T}),
%\end{align}
%and
%\begin{align}
%C_H=S_R\log_2(1+\frac{\eta^2\epsilon_r}{L}\Lambda^*)+S_T\log_2(1+\frac{(1-\eta^2)\epsilon_t}{L}\frac{1-S_R\Lambda^*}{S_T}),
%\end{align}
%respectively.
%\begin{proof}
%	See Appendix~\ref{app_rate}.
%\end{proof}

\vspace{-0.7cm}
\section{Type Selection for the RIS}
%\vspace{-0.2cm}
\label{moe_sele}

\vspace{-0.1cm}
\subsection{Reflective versus transmissive}
\vspace{-0.1cm}
Denote the function between the sum-rate with type $X$ ($X$ could be $T$, $R$, or $H$) and number $S_T$ of UEs within the transmission zone by $C_X(S_T)$. To compare the reflective type with the transmissive type, we first show how $C_R$ and $C_T$ change with $S_T$ in the following proposition. 
\vspace{-0.2cm}
\begin{proposition}
	\label{remark_mono}
	When $S_T$ increases, sum-rate $C_T$ becomes larger while $C_R$ degrades.
\end{proposition}
\vspace{-0.3cm}
\begin{proof}
	To show that $C_T$ is positively correlated with $S_T$, we consider the derivative of $C_T(S_T)$ within $(1,S-1)$, i.e.,
	{\setlength\abovedisplayskip{0.05cm}
		\setlength\belowdisplayskip{0cm}
	\begin{align}
	\label{derivative_C_T}
		C_T'(S_T)\!=\!\frac{\left(1\!+\!\frac{\epsilon_t}{L}\frac{1}{S_T}\right)\left(\ln(2)\log_2\left(1\!+\!\frac{\epsilon_t}{L}\frac{1}{S_T}\right)\!-\!1\right)\!+\!1}{\ln(2)\left(1\!+\!\frac{\epsilon_t}{L}\frac{1}{S_T}\right)}.
	\end{align}
	Define $f(x)=x(\ln(2)\log_2(x)-1)+1$. It can be found that $f(x)>0$ when $x\in(1,\infty)$. Therefore, $C_T'(S_T)$ is always positive, i.e., $C_T$ increases with $S_T$.}
	
	Similarly, we can prove that $C_R$ decreases with $S_T$. Due to the space limit, the proof is neglected here. 
\end{proof}
\vspace{-0.2cm}
%Define $g(x)=\log_2(1+\frac{\epsilon_t}{L}\frac{1}{x})$. Therefore, $C_T=g(S_T)$. It is easy to prove that $g$ is strictly monotonically increasing over $[1,(S_R+S_T)-1]$. Therefore, $C_T$ increases with number $S_T$ of UEs satisfying $y<0$. Similarly, we can prove that $C_R$ decreases when $S_T$ becomes larger. 

Assume that power radiation pattern $F_{m,n}^{(s)}$ of the $(m,n)$-th RIS element is approximately isotropic, i.e., $\epsilon_t\approx\epsilon_r$. Therefore, we have $C_T(1)-C_R(1)\le 0$ while $C_T(S-1)-C_R(S-1)\ge 0$. As a result, according to Proposition~\ref{remark_mono}, the UE distribution leading to the same performance for the reflective type and transmissive type is unique, which is denoted by $S_T^{(c)}$, i.e., $C_T(S_T^{(c)})=C_R(S_T^{(c)})$. Based on the monotonicity of $C_T$ and $C_R$, we can conclude that \emph{the transmissive type outperforms the reflective type when $S_T>S_T^{(c)}$, while the reflective type is better when $S_T\le S_T^{(c)}$}. 

To derive the optimal type, we will compare the hybrid type with one of the reflective and transmissive types that achieves a higher sum-rate in the following. Based on above analysis, the discussion is naturally divided into two parts, i.e., $S_T\le S_T^{(c)}$ and $S_T>S_T^{(c)}$, where the reflective type and transmissive type are compared with the hybrid type, respectively. 
\vspace{-0.4cm}
\subsection{Hybrid versus reflective}
Before comparison, we simplify the expression of $C_H$ first. Since $\epsilon_t\approx\epsilon_r$, we have $L\frac{2S_T}{S}\left(\frac{1}{\epsilon_t}-\frac{1}{\epsilon_r}\right)+\frac{1}{S}\approx \frac{1}{S}$. Note that $\frac{1}{S}$ is upper and lower bounded by $\frac{1}{S_R}$ and $0$, respectively. Therefore, $\Lambda^*$ in the expression of $C_H$ can be simplified into 
{\setlength\abovedisplayskip{0cm}
	\setlength\belowdisplayskip{0.1cm}
	\begin{align}
	\label{Lambda_simplify}
	\Lambda^*=L\frac{2S_T}{S}\left(\frac{1}{\epsilon_t}-\frac{1}{\epsilon_r}\right)+\frac{1}{S},
	\end{align}

To compare the hybrid type with the reflective type, we will find out the trend of $C_R-C_H$ with respect to $S_T$ in the following.} Assume that the transmit SNR $\frac{P_T}{\sigma^2}$ is sufficiently high such that the received SNR for each UE with the hybrid type RIS is sufficiently high, i.e., $\frac{P_T}{\sigma^2}\Lambda_s\hat{\beta}^{(s)}K_tMN|\Gamma_{m,n}^{(s)}|^2\gg 1$. Then, we have the following remark,
\vspace{-0.2cm}
\begin{remark}
	\label{remark_larger}
	The derivative of $C_R(S_T)$ is much larger than that of $C_H(S_T)$, i.e., $|C_R'(S_T)|\gg|C_H'(S_T)|$.
\end{remark}
\vspace{-0.3cm}
\begin{proof}
	See Appendix~\ref{app_larger}.
\end{proof}
\vspace{-0.2cm}
Therefore, $C_R'(S_T)-C_H'(S_T)$ can be approximated by $C_R'(S_T)$. According to Proposition~\ref{remark_mono}, $C_R$ is negatively correlated with $S_T$, and thus, $C_R-C_H$ also decreases with $S_T$. Define $C_{eq}$ as the sum-rate with the reflective (transmissive) type RIS 
%%%%%%%%%%%%%%%%%%%%%%%%%%%%%%%%%%%%%%%%%%%%%%%%%
\begin{table*}[!htbp]
	\centering
	\vspace{-0.5cm}
	\caption{\normalsize{Optimality conditions for RIS types}}
	\label{opt_condition}
	\vspace{-0.2cm}
	\begin{tabular}{|c|c|c|c|} %表格7列 全部居中显示
		\hline
		\multicolumn{3}{|c|}{\textbf{Conditions}}&\multirow{2}*{\tabincell{c}{\textbf{Optimal}\\ \textbf{type}}}\\
		\cline{1-3}
		\tabincell{c}{Sum rate comparison for hybrid and\\ reflective types when $S_T=S_T^{(c)}$} & \tabincell{c}{Sum rate comparison for hybrid and reflective types when $S_T=1$, \\and that for hybrid and transmissive types when $S_T=S-1$} & \tabincell{c}{Number of UEs within\\ the transmission zone}&\\
		\hline
		\hline
		\multirow{2}*{$C_H(S_T^{(c)})\le C_{eq}$}&\diagbox{ }{ }&$S_T>S_T^{(c)}$&T\\  %纵向合并4行单元格 
		\cline{2-4}  %为第二列到第七列添加横线
		&\diagbox{ }{ }&$S_T\le S_T^{(c)}$&R\\
		
		\hline
		\multirow{8}*{$C_H(S_T^{(c)})> C_{eq}$}&$C_H(1) > C_R(1)$, and $C_H(S-1) > C_T(S-1)$ & \diagbox{ }{ } & H\\
		\cline{2-4}
		&\multirow{2}*{$C_H(1) > C_R(1)$, and $C_H(S-1) \le C_T(S-1)$} & $S_T\ge S_T^{(b)}$& T\\
		\cline{3-4}
		&& $S_T< S_T^{(b)}$& H \\
		\cline{2-4}
		&\multirow{2}*{$C_H(1) \le C_R(1)$, and $C_H(S-1) > C_T(S-1)$} & $S_T\le S_T^{(a)}$ & R \\
		\cline{3-4}
		&& $S_T> S_T^{(a)}$ & H\\
		\cline{2-4}
		& \multirow{3}*{$C_H(1) \le C_R(1)$, and $C_H(S-1) \le C_T(S-1)$} & $S_T\le S_T^{(a)}$ & R\\
		\cline{3-4}
		&& $S_T\in(S_T^{(a)},S_T^{(b)})$& H\\
		\cline{3-4}
		&& $S_T\ge S_T^{(b)}$ & T\\
		\hline
		
	\end{tabular}
	\vspace{-0.4cm}
\end{table*}
%%%%%%%%%%%%%%%%%%%%%%%%%%%%%%%%%%%%%%%%%%%%%%%%
when $S_T^{(c)}$ UEs are located within the transmission zone, i.e., $C_{eq}\triangleq C_R(S_T^{(c)})=C_T(S_T^{(c)})$. Due to the monotonicity, the values of $C_R-C_H$ at $S_T=1$ and $S_T=S_T^{(c)}$ have an influence on the performance of the reflective type and hybrid type. Specifically, 1) when $C_R(1)-C_H(1)\ge 0$ and $C_{eq}-C_H(S_T^{(c)})\ge0$, the reflective type is better than the hybrid type. 2) When $C_R(1)-C_H(1)< 0$ and $C_{eq}-C_H(S_T^{(c)})<0$, the hybrid type outperforms the reflective type. 3) When $C_R(1)-C_H(1)\ge 0$ and $C_{eq}-C_H(S_T^{(c)})<0$, the reflective type is better if $S_T\in[1,S_T^{(a)}]$, while the hybrid type achieves a larger sum-rate if $S_T\in(S_T^{(a)},S_T^{(c)}]$. Here, $S_T^{(a)}$ is the unique UE distribution that leads to the same performance for the reflective and hybrid types.
%\begin{enumerate}[itemindent=0em, label=$\bullet$]
%	\item When $C_R(1)-C_H(1)\ge 0$ and $C_{eq}-C_H(S_T^{(c)})\ge0$, the reflective type is better than the hybrid type.
%	\item When $C_R(1)-C_H(1)< 0$ and $C_{eq}-C_H(S_T^{(c)})<0$, the hybrid type outperforms the reflective type.
%	\item When $C_R(1)-C_H(1)\ge 0$ and $C_{eq}-C_H(S_T^{(c)})<0$, the reflective type is better if $S_T\in[1,S_T^{(a)}]$, while the hybrid type achieves a larger sum-rate if $S_T\in(S_T^{(a)},S_T^{(c)}]$. Here, $S_T^{(R,H)}$ represents the unique UE distribution that leads to the same performance for the reflective and hybrid types.
%\end{enumerate}
It is worthwhile noting that when $C_R(1)-C_H(1)< 0$, $C_{eq}-C_H(S_T^{(c)})$ cannot take positive values, since $C_R-C_H$ decreases with $S_T$.
\vspace{-0.4cm}
\subsection{Hybrid versus transmissive}
Similar to Remark~\ref{remark_larger}, it can be proved that $|C_T'(S_T)|$ is much larger than $|C_H'(S_T)|$, and thus, $C_T-C_H$ increases with $S_T$ according to Proposition~\ref{remark_mono}. Therefore, the values of $C_T-C_H$ at $S_T=S_T^{(c)}$ and $S_T=S-1$ also have an effect on the performance of the hybrid and transmissive types, where the detailed discussion is omitted here due to the space limit. Define $S_T^{(b)}$ as the unique UE distribution that leads to the same performance for the transmissive and hybrid types.
%Specifically, it can be obtained that \textbf{when} $C_T(1)-C_H(1)\ge 0$ and $C_{eq}-C_H(S_T^{(c)})\ge0$, the transmissive type outperforms the hybrid type. \textbf{When} $C_T(1)-C_H(1)< 0$ and $C_{eq}-C_H(S_T^{(c)})<0$, the hybrid type is preferred. \textbf{When} $C_T(1)-C_H(1)\ge 0$ and $C_{eq}-C_H(S_T^{(c)})<0$, the transmissive type is better if $S_T\in[S_T^{(b)},S-1]$, while the hybrid type achieves a larger sum-rate if $S_T\in (S_T^{(c)},S_T^{(b)})$. 
Based on above discussions, the optimal conditions for each type can be summarized in Table~\ref{opt_condition}, where $T$, $R$, and $H$ represent transmissive, reflective, and hybrid types, respectively.

It can be found that the number $(M,N)$ of RIS elements, distance $D_c^{(c)}$ between the BS and the RIS, and distance $d_{c,U}$ between the RIS and UEs will influence the relationships of $\left(C_H(1),C_R(1)\right)$, $\left(C_H(S-1), C_T(S-1)\right)$, and $\left(C_{eq},C_H(S_T^{(c)})\right)$, which can change the optimal type. Therefore, we will discuss the effect of $(M,N)$ and $(D_c^{(c)},d_{c,U})$ on the optimal type in the following.
\vspace{-0.2cm}
\begin{proposition}
	\label{prop_dist}	
	Given the number of the RIS elements, when both the UEs and the BS are far from the RIS, the hybrid type is always inferior to the reflective or transmissive type.
\end{proposition}
\vspace{-0.3cm}
\begin{proof}
	See Appendix~\ref{app_dist}.
\end{proof}
\vspace{-0.3cm}
\begin{proposition}
	\label{prop_MN}
	When the number of RIS elements is sufficiently large, i.e., $MN>E_02^{\max(E_R, E_T)}$,
%	\begin{align}
%	MN>E_02^{\max(E_R, E_T)},
%	\end{align}
	the hybrid type is always better than the other two types for an arbitrary UE distribution. Here,
	{\setlength\abovedisplayskip{0cm}
		\setlength\belowdisplayskip{0.1cm}
	\begin{gather}
%	E_0=\frac{64\pi^3(D_{c}^{(c)}d_{c,U})^{\alpha}}{\gamma_0\lambda^2Gl_Ml_NG_{I}\cos^2\theta_c^{(c)}K_t},\notag\\
E_0=64\pi^3(D_{c}^{(c)}d_{c,U})^{\alpha}\sigma^2/(P_T\lambda^2Gl_Ml_NG_{I}\cos^2\theta_c^{(c)}K_t),\notag\\
%	E_R=\frac{2S\log_2\Gamma_H^r+S_T\log_2{\epsilon_r}-S\log_2{S}+S_R\log_2{S_R}}{-S_T},\notag\\
	E_R\!=\!(S\!-\!S_T\log_2{\epsilon_r}\!+\!S\log_2{S}\!-\!S_R\log_2{S_R})/S_T,\notag\\
%	E_T=\frac{2S\log_2\Gamma_H^r+S_R\log_2{\epsilon_t}-S\log_2{S}+S_T\log_2{S_T}}{-S_R}.\notag
	E_T\!=\!(S\!-\!S_R\log_2{\epsilon_t}\!+\!S\log_2{S}\!-\!S_T\log_2{S_T})/S_R.\notag
	\end{gather}}
\end{proposition}
\vspace{-0.2cm}
\begin{proof}
	As shown in Appendix~\ref{app_dist}, we have 
	{\setlength\abovedisplayskip{0cm}
		\setlength\belowdisplayskip{0.05cm}
	\begin{align}
	C_H-C_T\approx& -S+S_R\log_2{\epsilon_t}-S_R\log_2{L}\notag\\
	&-\left(S\log_2{S}-S_T\log_2{S_T}\right).
	\end{align}
	By substituting the expression of $L$, it can be obtained that the hybrid type outperforms the transmissive type only when RIS size $MN$ satisfies $MN>E_02^{E_T}$. Similarly, we can prove that the hybrid type achieves a higher sum rate than the reflective type only when $MN$ satisfies $MN>E_02^{E_R}$, which ends the proof.}
\end{proof}

\vspace{-0.8cm}
\section{Simulation Results}
\vspace{-0.2cm}
\label{simu}
In this section, we verify the effectiveness of the derivation and the proposed optimality conditions by simulations. The parameters are based on 3GPP documents~\cite{3GPP_38803} and existing works~\cite{BHLYZH_2020,BLY_2016}. Specifically, the RIS is placed vertical to the ground, with one of its edge parallel to the ground. Besides, the RIS is deployed vertical to the direction from the BS to the RIS. The height of the BS antennas and RIS center are set as $30$~m and $15$~m, respectively. The BS is equipped with $K_t=12$ antennas with a transmit power of $P_T=43$~dBm. The wavelength is set as $\lambda=0.1$~m and the noise variance is $\sigma^2=-96$~dBm. The antenna gain and pathloss exponent are given by $G=1$ and $\alpha=2$, respectively. We set the size of one RIS element as $l_M\!=\!l_N\!=\!0.02$~m. Besides, power radiation parameters of one element are given by $\epsilon_r\!=\!1$ and $\epsilon_t\!=\!0.95$, with the antenna gain of one element $G_I\!=\!1$. 

\begin{figure*}[!tpb]
	\centering
	\subfigure[]{
	\begin{minipage}[b]{0.28\textwidth}
		\centering
		\includegraphics[width=1\textwidth]{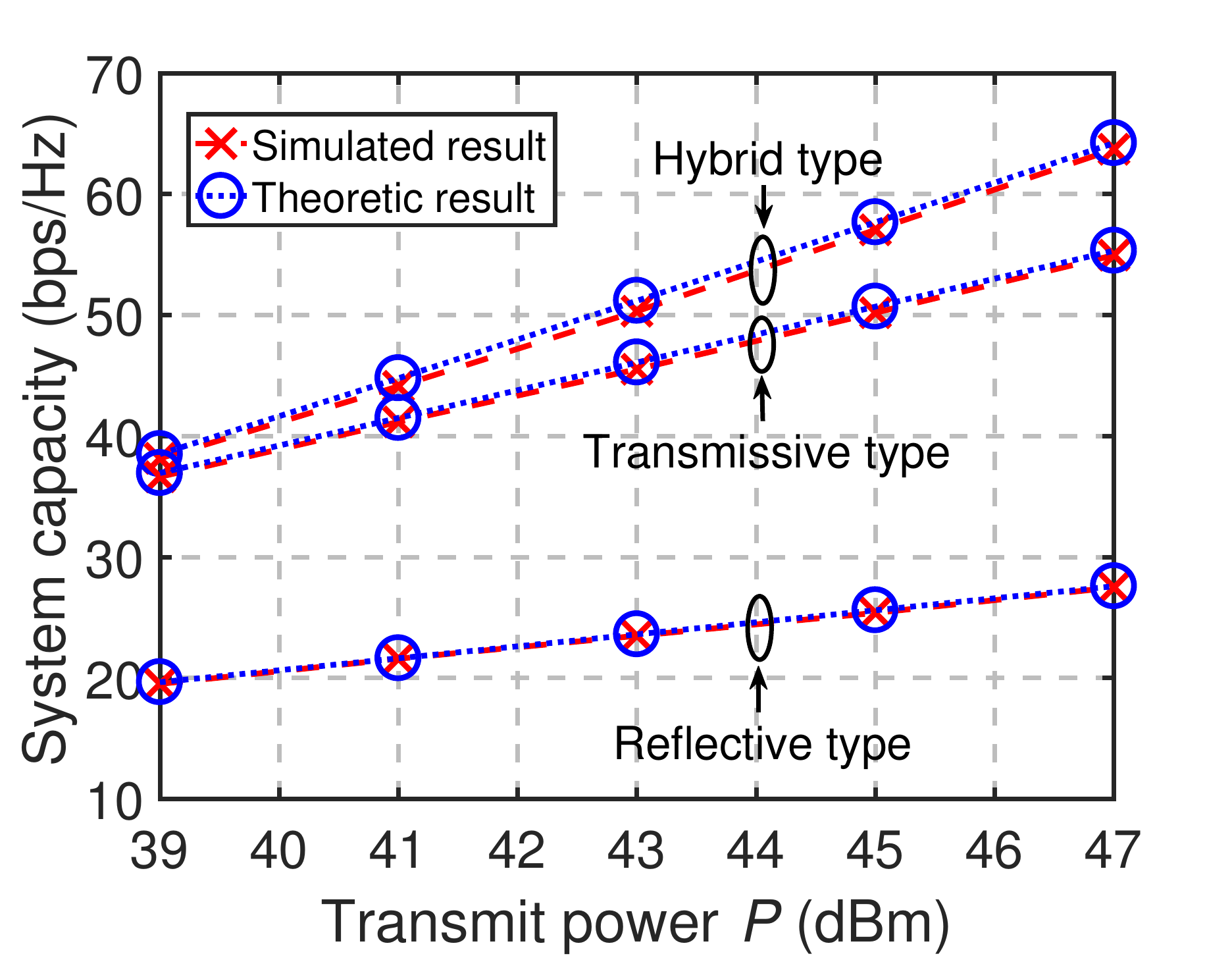}
		\vspace{-0.5cm}
		\label{simu_P}
	\end{minipage}}
	\subfigure[]{
	\begin{minipage}[b]{0.28\textwidth}
		\centering
		\includegraphics[width=1\textwidth]{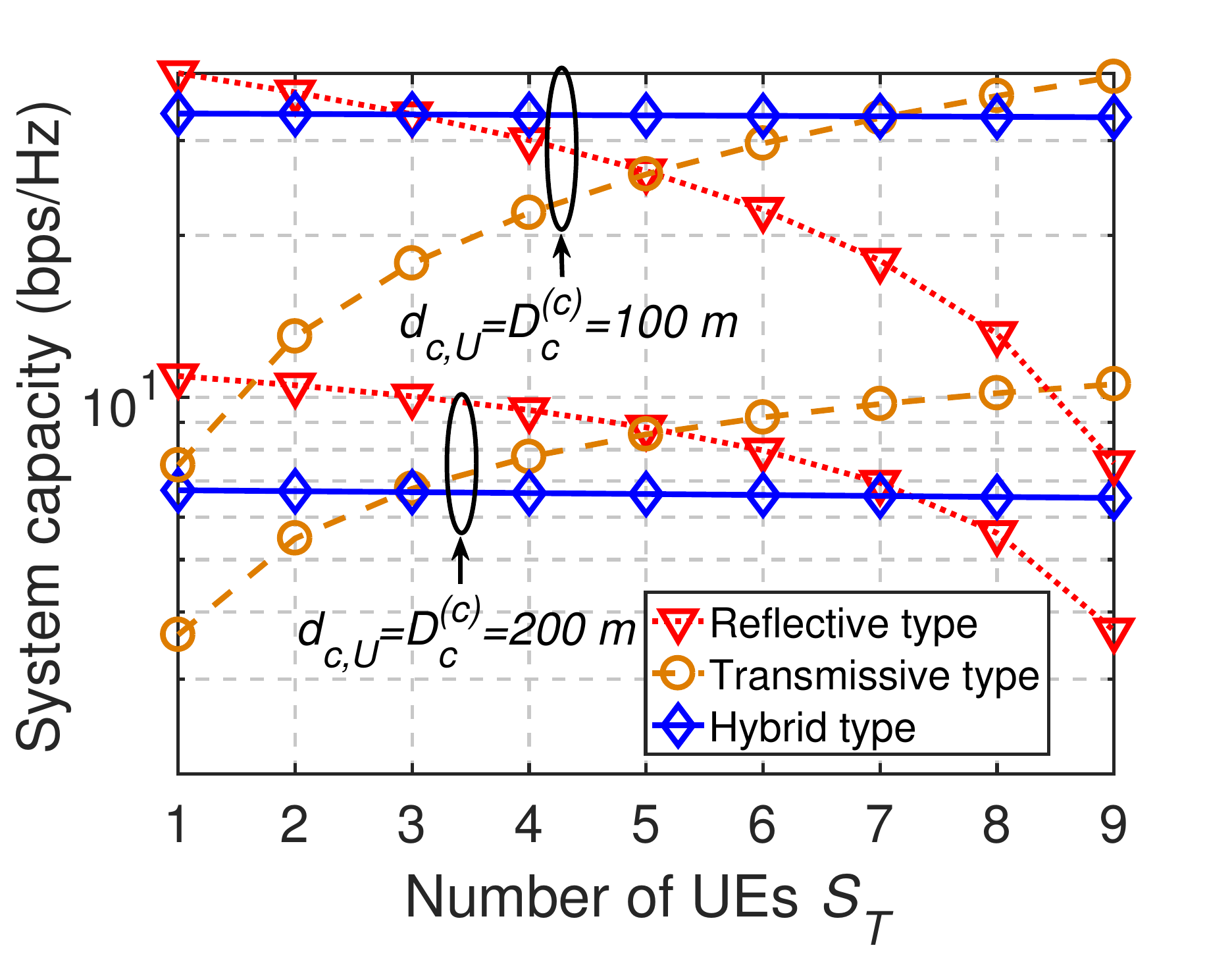}
		\vspace{-0.6cm}
		%			\vspace{-0.3cm}
		\label{simu_M_100}
	\end{minipage}}
	\subfigure[]{
	\begin{minipage}[b]{0.28\textwidth}
		\centering
		\includegraphics[width=1\textwidth]{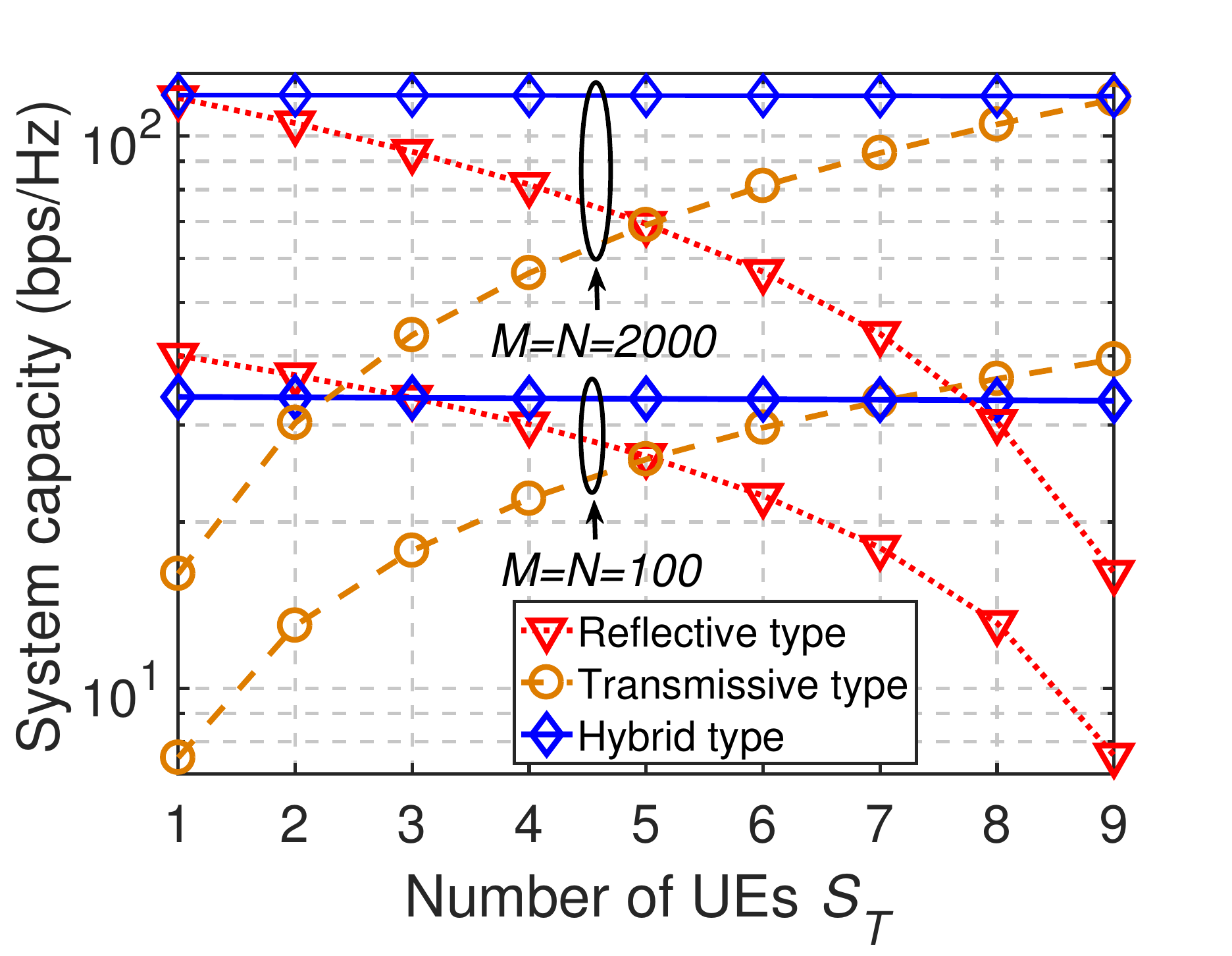}
		\vspace{-0.6cm}
		%			\vspace{-0.3cm}
		\label{simu_d_50}
	\end{minipage}}
\vspace{-0.4cm}
\caption{Simulation results: (a) System capacity vs. transmit power with $S_T\!=\!7$, $S=10$, $M=N=50$, and $d_{c,U}\!=\!D_c^{(c)}=\!50$~m. (b) System capacity vs. number of UEs within the transmission zone $S_T$, with $M=N=100$ and $S=10$. (c) System capacity vs. number of UEs within the transmission zone $S_T$, with $d_{c,U}\!=\!D_c^{(c)}\!=\!100$~m and $S=10$.}
	\vspace{-0.7cm}
\end{figure*}

%\begin{figure*}[!tpb]
%	\centering
%%	\vspace{-0.2cm}
%		\begin{minipage}[b]{0.3\textwidth}
%			\centering
%			\includegraphics[width=0.8\textwidth]{sum_rate_vs_transmit_power.eps}
%			\vspace{-0.4cm}
%			\caption{Sum rate vs. transmit power with \newline $S_T=7$, $M=N=50$, and $d_{c,U}\!=\!D_c^{(c)}$\newline$=\!50$~m.}
%%			\vspace{-0.3cm}
%			\label{simu_P}
%	\end{minipage}
%		\begin{minipage}[b]{0.3\textwidth}
%			\centering
%			\includegraphics[width=0.8\textwidth]{sumrate_vs_S_T_M=N=100.eps}
%			\vspace{-0.4cm}
%			\caption{Sum rate vs. number of UEs $S_T$, \newline with $M=N=100$.}
%%			\vspace{-0.3cm}
%			\label{simu_M_100}
%	\end{minipage}
%		\begin{minipage}[b]{0.3\textwidth}
%			\centering
%			\includegraphics[width=0.8\textwidth]{sumrate_vs_S_T_d=50.eps}
%			\vspace{-0.4cm}
%			\caption{Sum rate vs. number of UEs $S_T$, \newline with $d_{c,U}\!=\!D_c^{(c)}\!=\!50$~m.}
%%			\vspace{-0.3cm}
%			\label{simu_d_50}
%	\end{minipage}
%\vspace{-0.7cm}
%\end{figure*}

Fig.~\ref{simu_P} shows the system capacity versus transmit power of the BS. The simulated and theoretic results correspond to the data rate $C$ and its upper bound $C^{ub}$, respectively, where the simulated results is obtained by $100$ Monte Carlo simulations. From Fig.~\ref{simu_P}, we can find that $C$ is upper bounded by $C^{ub}$ for the three RIS types, which verifies Proposition~\ref{prop_cap_ub}. Besides, according to Fig.~\ref{simu_P}, the gap between the capacity and its upper bound is small, which justifies the selection of $C^{ub}$ as the performance metric for RIS types.

Fig.~\ref{simu_M_100} depicts the system capacity versus number of UEs within the transmission zone $S_T$, under different distances $\big(d_{c,U},D_c^{(c)}\big)$. We can find that despite the values of $\big(d_{c,U},D_c^{(c)}\big)$, the capacity in the reflective type always decreases with $S_T$ while the sum-rate achieved by the transmissive type is positively correlated with $S_T$, which is consistent with Proposition~\ref{remark_mono}. Besides, from Fig.~\ref{simu_M_100}, it can be observed that when both distances are set as $200$~m, the hybrid type is always inferior to the reflective or transmissive type, which consists with Proposition~\ref{prop_dist}. According to Fig.~\ref{simu_M_100}, we can also find that when both the UEs and the BS are closer to the RIS, the advantage of the hybrid type over the other two types becomes more significant.
%since more UEs are scheduled when the hybrid type is deployed.

Fig.~\ref{simu_d_50} shows the system capacity versus $S_T$, under different RIS sizes $(M,N)$. From Fig.~\ref{simu_d_50}, we can observe that when the RIS contains more elements, the advantage of the hybrid type over the other two types is more significant, since more UEs are scheduled when the hybrid type RIS is adopted. In addition, according to Fig.~\ref{simu_d_50}, it can be found that when the number of RIS elements is sufficiently large, the hybrid type always outperforms the other two  types for arbitrary $S_T$, which is consistent with Proposition~\ref{prop_MN}.

%\begin{figure}[!tpb]
%	\centering
%	\includegraphics[width=2.5in]{sum_rate_vs_UE_num.eps}
%%	\vspace{-0.4cm}
%	\caption{Sum rate vs. ratio of UEs on two sides of the RIS, with $S_T=10$.}
%%	\vspace{-0.6cm}
%	%	\setlength{\belowcaptionskip}{-1cm}
%	\label{s_vs_n}
%\end{figure}
%
%
%
%\begin{figure}[!tpb]
%	\centering
%	\includegraphics[width=2.5in]{sum_rate_vs_Eratio.eps}
%	%	\vspace{-0.4cm}
%	\caption{Sum rate vs. $\frac{E_T}{E_R}$, with $S_T=10$, and $S_R=20$.}
%	%	\vspace{-0.6cm}
%	%	\setlength{\belowcaptionskip}{-1cm}
%	\label{s_vs_r}
%\end{figure}
%
%\begin{figure}[!tpb]
%	\centering
%	\includegraphics[width=2.5in]{sum_rate_vs_Eratio_b.eps}
%	%	\vspace{-0.4cm}
%	\caption{Sum rate vs. $\frac{E_T}{E_R}$, with $S_T=30$, and $S_R=20$.}
%	%	\vspace{-0.6cm}
%	%	\setlength{\belowcaptionskip}{-1cm}
%	\label{s_vs_r_b}
%\end{figure}

\vspace{-0.5cm}
\section{Conclusion}
\vspace{-0.2cm}
In this letter, we have considered a downlink RIS-assisted network with one BS and multiple UEs. The system capacity of this network has been derived, based on which the optimal type of the RIS under a specific UE distribution has been analyzed. From the analysis and simulation, we can conclude that: 1) When both the BS and the UEs are far away from the RIS, the hybrid type is always inferior to the reflective or transmissive type. 2) When the number of RIS elements exceeds the derived threshold, the hybrid type is always better than the other types for an arbitrary UE distribution.
\label{conclu}
\begin{appendices}
\vspace{-0.6cm}
\section{Proof of Remark~\ref{remark_larger}}
\vspace{-0.2cm}
\label{app_larger}
The derivative of $C_H$ can be given by
{\setlength\abovedisplayskip{0cm}
	\setlength\belowdisplayskip{0cm}
%\begin{footnotesize}
\begin{align}
&C_H'(S_T)\!=\!\Big(\!-\!\log_2\left(1\!+\!\frac{\epsilon_r}{2L}\Lambda^*\!\right)\!+\!\log_2\Big(\!1\!+\!\frac{\epsilon_t}{2L}\frac{1\!-\!S_R\Lambda^*}{S_T}\Big)\!\Big)\notag\\
&\!+\!\bigg(\frac{S_R}{S}\frac{A_1}{\ln{2}\left(\!1\!+\!\frac{\epsilon_r}{2L}\Lambda^*\!\right)}\!+\!\frac{S_T}{S}\frac{A_2}{1\!+\!\frac{\epsilon_t}{2L}\frac{1\!-\!S_R\Lambda^*}{S_T}}\!\bigg) \!\triangleq\! E_1\!+\!E_2, \notag
\end{align}
%\end{footnotesize}
where $A_1=\frac{\epsilon_r}{\epsilon_t}-1$ and $A_2=1-\frac{\epsilon_t}{\epsilon_r}$. In the following, we will show $E_2\approx 0$ and $E_1\approx 0$.}

 Since $\epsilon_t\approx\epsilon_r$, we have $A_1\approx 0$ and $A_2\approx 0$. Besides, since $\frac{S_T}{S}<1$, $\frac{S_R}{S}<1$, and the received SNR at each UE with the hybrid type is sufficiently large, we can obtain that $E_2\approx 0$.  
 
 By substituting the expression of $\Lambda^*$ in (\ref{Lambda_simplify}), we have $1+\frac{\epsilon_r}{2L}\Lambda^*=1+\frac{\epsilon_r}{2LS}+\frac{S_T}{S}A_1$.
% \begin{align}
% 1+\frac{(\Gamma_H^r)^2\epsilon_r}{L}\Lambda^*=1+\frac{(\Gamma_H^r)^2\epsilon_r}{LS}+\frac{S_T}{S}A_1.
% \end{align} 
 Since $1+\frac{\epsilon_r}{2L}\Lambda^*\gg 1$ while $\frac{S_T}{S}A_1\approx 0$, we can obtain $\log_2\left(1+\frac{\epsilon_r}{2L}\Lambda^*\right)\approx \log_2\left(\frac{\epsilon_r}{2LS}\right).$
% \begin{align}
% \log_2(1+\frac{(\Gamma_H^r)^2\epsilon_r}{L}\Lambda^*)\approx \log_2(\frac{(\Gamma_H^r)^2\epsilon_r}{LS}).
% \end{align}
 Similarly, it can be obtained that $\log_2\left(1+\frac{\epsilon_t}{2L}\frac{1-S_R\Lambda^*}{S_T}\right)\approx \log_2\left(\frac{\epsilon_t}{2LS}\right)$
% \begin{align}
% \log_2(1+\frac{(1-(\Gamma_H^r)^2)\epsilon_t}{L}\frac{1-S_R\Lambda^*}{S_T})\approx \log_2(\frac{(1-(\Gamma_H^r)^2)\epsilon_t}{LS}).\notag
% \end{align}
 Therefore, $E_1$ can be rewritten as $E_1=\log_2\left(\frac{\epsilon_t}{\epsilon_r}\right)$,
% \begin{align}
% E_1=\log_2(\frac{(1-\eta^2)\epsilon_t}{\eta^2\epsilon_r}),
% \end{align}
 from which we have $E_1\approx 0$. Therefore, $C_H'(S_T)\approx 0$.
 
 According to (\ref{derivative_C_T}), it can be found that the value of $C_T'(S_T)$ is large, which ends the proof.
% On the contrary, the derivative of $C_R$ can be expressed as
% \begin{align}
% C_R'(S_T)=
% \end{align}
\vspace{-0.4cm}
\section{Proof of Proposition~\ref{prop_dist}}
\vspace{-0.2cm}
\label{app_dist}
We first show that $C_H$ is smaller than $C_T$ for any given $S_T$. Specifically, we have
{\setlength\abovedisplayskip{0cm}
	\setlength\belowdisplayskip{0cm}
%	\begin{footnotesize}
\begin{align}
\label{difference}
&C_H-C_T\approx S\log_2\left(\frac{\epsilon_t}{2LS}\right)-S_T\log_2\left(\frac{\epsilon_t}{LS_T}\right)\notag\\
\!=&\!-\!S\!+\!S_R\log_2{\epsilon_t}\!-\!S\log_2{S}\!+\!S_T\log_2{S_T}\!-\!S_R\log_2{L}.
\end{align}
%\end{footnotesize}
%}
Since} both the UEs and the BS are far away from the RIS, $L$ takes a large value. Besides, the other terms in (\ref{difference}) are not influenced by the distance between the UEs and the RIS, and the distance between the BS and the RIS. Therefore, we have $C_H-C_T< 0$, and thus, the hybrid type is inferior to the transmissive type, which ends the proof.

\end{appendices}

\vspace{-0.4cm}

%%%%%%%%%%%%%%%%%%%%%%%%%%%%%%%%%%%%%%%%%%%%%%%

\end{document}